\documentclass[aps,prc,twocolumn,superscriptaddress,showpacs,floatfix]{revtex4}
\usepackage{graphicx}
\usepackage[german,USenglish]{babel}

\begin{document}

\title{Spectra and binding energy predictions of chiral interactions for $^7$Li}

\author{A. Nogga}
\affiliation{Institut f\"ur Kernphysik, Forschungszentrum J\"ulich, 52425 J\"ulich, Germany}
\email{a.nogga@fz-juelich.de}

\author{P. Navr{\'a}til}
\affiliation{Lawrence Livermore National Laboratory, L-14, P.O. Box 808, Livermore, 
         California 94551, USA}
\email{navratil1@llnl.gov}

\author{B.R. Barrett}
\affiliation{Department of  Physics, University of Arizona, Tucson, Arizona 85721, USA}
\email{bbarrett@physics.arizona.edu}

\author{J.P. Vary}
\affiliation{Department of Physics and Astronomy, Iowa State University, Ames, Iowa 50011, USA}
\email{jvary@iastate.edu}

\date{November 28, 2005}

\begin{abstract}
Using the no-core shell model approach, 
we report on the first results for
$^{7}$Li  based on the next-to-next-to-leading order 
chiral nuclear interaction. Both, two-nucleon and three-nucleon interactions 
are taken into account. We show that the $p$-shell nuclei are sensitive to the subleading
parts of the chiral interactions including three-nucleon forces. Though chiral 
interactions are soft, we do not observe overbinding for this $p$-shell nucleus and 
find a realistic description for the binding energy, excitation spectrum and radius.   
\end{abstract}


\maketitle

\section{Introduction}

We are currently experiencing rapid progress in our understanding of nuclear properties.
This is triggered by two major developments. 

On one hand, we have increasingly powerful supercomputers, for which new and very 
efficient algorithms have been developed to solve the nuclear many-body problem. 
We are now able to solve the Schr\"odinger equation for realistic nuclear interactions 
for $p$-shell nuclei including also three-nucleon forces (3NF's) \cite{pieper01b,navratil03}. 
This is a major advance in itself, because it is becoming 
more clear that reliable predictions for many nuclear observables, binding energy and spectra 
can be obtained from phenomenological nuclear two- (NN) and three-nucleon (3N) interaction 
models. Especially, for the NN system, these have reached a high degree of sophistication and 
describe the NN data up to pion production threshold perfectly 
\cite{stoks94,wiringa95,machleidt96,machleidt01a}. For 
the binding energies of $p$-shell nuclei, the structure of the 3NF's turned out to be 
significant, leading to improved models of these forces engineered to  
describe a wide range of light nuclei accurately \cite{pieper01a}. 
This tool is of increasing importance to determine, e.g. reaction rates for 
astrophysical  processes \cite{park03}, which are experimentally not accessible or properties of  nuclei 
with large neutron excess \cite{pieper01b}. 

On the other hand,  there was a great deal of progress in our understanding of how 
chiral perturbation theory (ChPT)  can be extended from purely 
pionic or the nucleon-pion system (for a review see e.g. \cite{bernard95})
to systems with more than one nucleon \cite{kolck99,bedaque02a,epelbaum05b}. 
In this approach, one makes use of the explicit and spontaneous breaking of chiral symmetry to systematically expand the strong interaction in terms of 
a generic small momentum. Thereby, the NN interactions,
the 3NF's and also $\pi$N scattering are related to each other. 
The chiral symmetry and the pattern of its breaking are not systematically 
taken into account by today's phenomenological interactions, except 
that in all of them the longest range part of the potential  
is generally the one pion exchange interaction.
Therefore, though QCD is believed to be the theory of the strong interaction for 
the energies of relevance in nuclear systems, we are not able to perform 
confirming tests using the traditional forces.
This will be possible using nuclear forces based on ChPT. It will be especially 
important to look at subleading parts of the interaction, which include 
the 2$\pi$ exchange NN and 3N forces. Here, many of the 
relations between NN, 3N and $\pi$N interactions become apparent.
Therefore, finding signatures of the 3NF's is an important
aspect of current research on this issue. 

In the past, the effects of 3NF's have been studied using the phenomenological models.
In nucleon-deuteron (Nd) scattering above 60-100~MeV lab energy, it was found that predictions for some 
(polarization) observables depend on the 3NF model used, but not on the NN interaction 
chosen \cite{witala01a}.  These observables are excellent candidates to pin down 
the structure of 3NF's and, therefore, outstanding laboratories to study the important 
subleading parts of the chiral interactions. Consequently, a lot of experiments were triggered, which provide an important set of data to probe these models 
\cite{bieber00a,sakai00,cadman01,ermisch01}. First comparisons for chiral 
interactions to the data were performed and, in general, agreement was found 
in the energy range, where chiral interactions are expected to work \cite{epelbaum02c}.

These data, however, are manifestly isospin $T=1/2$ and, therefore, not suitable to 
probe the 3NF in the isospin $T=3/2$ channel. But it was argued that the $T=3/2$ component is very important to describe $p$-shell nuclei, which mainly motivated 
the new 3NF terms in the Illinois series of 3N interactions \cite{pieper01a}. 
This clearly shows that the spectra and binding energies of light nuclei are 
complementary to nd scattering and provide indispensable information on 
the structure of 3NF's. It is therefore of utmost importance to predict properties
of $p$-shell nuclei based on the chiral interactions including these 3NF's. 

Chiral interactions are low momentum interactions. The unknown short 
distance part of the force is absorbed in a tower of contact terms. 
Necessarily, we need to regularize the interactions with cutoff 
functions. It turns out that one obtains a decent description of 
the NN data using rather small cutoffs. This is advantageous, because
the resulting interactions are soft and convergence is generally faster
than the convergence with traditional models. But this has also been criticized in the past.
The experience with traditional models indicated a need for 
rather hard cores not only in the NN interaction, but also in the 3NF to 
prevent strong overbinding in systems beyond the $s$-shell. 
This is a surprising result, because it means that we cannot 
properly separate high energy degrees of freedom 
from the low energy part. Because the high momentum tail of traditional 
interactions is strongly model dependent, a sensitivity to this tail will induce 
a strong model dependence of the results. Indeed, the binding energies of 
nuclei do depend on the chosen NN interaction model. It is believed that 
this model dependence is strongly reduced, once 3NF's consistent with the 
NN force are added (in fact it was proven that any NN interaction 
can be augmented by a 3NF so that the description of the 3N data is identical 
to that of a second phase-equivalent NN force \cite{polyzou90}, thereby removing 
the model dependence). 
Similarly, one can expect that a consistent combination of NN and 3N force 
will lead to a reasonable description of light nuclei even if the 
high momentum tail is completely missing. An important aspect of this work 
will be to confirm this expectation by an explicit calculation.  

Because of the strong non-locality of today's chiral interactions, to the best 
of our knowledge, there is only one reliable many-body technique to solve the 
Schr\"odinger equation for $p$-shell nuclei for these interactions including the 3NF's. 
This is the ab-initio no-core shell model (NCSM) \cite{navratil03,navratil00c,navratil00b}. 
In this paper, we use the NCSM to predict spectra and binding energies for $^7$Li.  These results will be presented along with calculations for the $s$-shell states 
for $^3$H and $^4$He, which were used to determine unknown parameters of the 3NF. 
The $p$-shell results are then predictions. 

The manuscript is organized as follows. In Section~\ref{sec:chiral}, we start with a brief 
summary of the status on the research on chiral interaction models, mainly to explain 
the force model entering in the calculations. The NCSM approach is introduced 
in Section~\ref{sec:ncsm} to keep this paper self-contained.  Then, in 
Section~\ref{sec:results}, we discuss our numerical results. 
Conclusions and an outlook will be given in Section~\ref{sec:concl}.

\section{Chiral interactions}
\label{sec:chiral}

Microscopic nuclear structure is generally based on NN interactions, which describe 
the NN scattering data up to $\pi$-production threshold perfectly. Usually, such
interactions are based on the 1$\pi$-exchange, which is augmented by a more or less
phenomenological short range part. It turns out that binding energies based on these 
interactions are model dependent and too small 
\cite{nogga02b,navratil98b,pieper01b}.  3NF's are clearly necessary. 

However, defining consistent combinations of NN and 3N forces is not a trivial task,
and rarely are both based on a single underlying theory. In practice, the NN 
interaction models are usually augmented by a 2$\pi$ exchange 3NF \cite{pudliner97,coon79} or extended versions thereof \cite{pieper01a}. Then 
parameters are mostly adjusted to reproduce the binding energies 
of $s$-shell nuclei or nuclear matter density. Though this can be very successful 
in the description of nuclear properties \cite{pieper01a}, it lacks a solid theoretical 
foundation. 
The relation to QCD is completely lost. Processes of other strongly interacting 
systems, like the $\pi$N system, are not quantitatively related to these forces, 
though the basic mechanisms involve these particles.  

Since lattice simulations for nuclei are not realistic for the foreseeable future, 
effective field theory is an appealing theoretical foundation on which to build. 
In its simplest form, an effective field theory was formulated, that 
explicitly takes only nucleons into account. It is accompanied by a power 
counting in terms of powers of $r/a$, where $r$ is the effective range of the 
interaction and $a$ the large scattering length of the NN system (for a recent review 
see \cite{braaten04}). This property is realized in very different physical systems,
e.g. $^4$He clusters and atoms close to a Feshbach resonance. It allows one
to identify universal properties. Also, it is possible to identify nuclear properties,
which are dominated by the large scattering length of the NN interaction. 
An example is the correlation between $^3$H and $^4$He binding energies, known 
as Tjon-line, which is naturally explained using this approach \cite{platter05}.   
While these results are interesting in themselves, they are by construction
not that interesting for the relation of QCD and nuclear physics. 
The approach will only describe observables, which do not reflect any 
specific property of QCD except the large scattering length. 

To sense the leading role of QCD,  
another effective field theory is the tool of choice: ChPT. 
The new physical input is the approximate chiral  symmetry of QCD, 
which is known to be spontaneously broken. Based on this an 
effective field theory can be formulated that involves nucleons and $\pi$'s
explicitly.  The Goldstone-boson character of the $\pi$'s (related to 
the spontaneous breaking of chiral symmetry) guarantees that amplitudes in purely 
pionic and the $\pi$N system can be expanded in powers of a small,
typical  momentum $Q$. 
The explicit breaking of chiral symmetry due to the finite 
quark masses is expanded similarly in terms of 
the $\pi$ mass \cite{weinberg79}. Electromagnetic effects on 
the quark level and quark mass differences  have also been taken into
account by an expansion of  isospin symmetry breaking contributions
in terms of the fine structure constant and the quark mass differences
\cite{fettes99,kolck96}. 

Obviously, the NN system is different. The deuteron bound state clearly shows 
that a perturbative treatment is impossible. Weinberg realized that this is 
understood by an enhancement of reducible diagrams in two- or more-nucleon 
systems and proposed to expand the irreducible diagrams perturbatively in
terms of small external momenta, but then sum all these diagrams to infinite 
order using a Lippmann-Schwinger or Schr\"odinger equation 
\cite{weinberg90,weinberg91}. In other words, Weinberg suggested to obtain 
nuclear potentials from ChPT.  This naturally explains that NN interactions 
are much more important than 3NF's and that these are more relevant than 
higher-body interactions \cite{weinberg92}.  The approach was quantitatively 
confirmed \cite{ordonez96,epelbaum00,entem02b}. For the NN system, terms 
up to order momentum $Q^4$ have been taken into account nowadays \cite{entem03a,epelbaum05a} and lead to a description of NN data, which 
is comparable to the one of the phenomenological interactions. The approach 
to sum all irreducible diagrams to infinite order is usually called ``Weinberg 
counting''. Other schemes to deal with the non-perturbativity have been proposed
\cite{kaplan98b} and are formally more consistent, because the 
renormalization is understood analytically. But the formal problems
of the ``Weinberg'' scheme, 
which were pointed out in \cite{kaplan98b}, can be circumvented  
by small cutoffs of the order of 500~MeV 
for the regularization of the Lippmann-Schwinger  equation. 
Under this condition,  ``Weinberg power-counting''
is numerically more successful than the scheme of \cite{kaplan98b} 
as was shown in \cite{fleming00}. Indirect confirmation of the Weinberg 
approach is the favorable agreement of  parameters extracted from 
$\pi$N scattering \cite{buttiker00} and an NN phase shift analysis \cite{rentmeester03,klomp91}. 
Therefore, for the rest of this paper,  we will rely on the Idaho-N3LO interaction \cite{entem03a} based on ``Weinberg counting'' for the NN interaction. It 
takes terms up to order $Q^4$ into account and is able 
to describe the NN scattering data below $\pi$-production threshold 
perfectly.  We stress that this can only be a first step, because a complete 
analysis of the chiral interactions requires further order-by-order analysis to 
confirm convergence. 

Very soon the leading 3NF's, which appear at order $Q^3$  were 
derived \cite{kolck94}. They consist of a 2$\pi$ 
exchange term, which is very similar to the widely used Tucson-Melbourne 
3NF \cite{coon79} and, additionally, there are contact interactions with and without 
1$\pi$ exchange. In the original paper, five such contact terms were believed 
to be independent. Fortunately, it turned out later, that only two of them remain 
independent once the identity of the nucleons is taken into account. In this simplified 
form the 3NF's were applied to Nd scattering, the $^3$H and $^4$He 
bound states \cite{epelbaum02c}. For  completeness, the explicit expressions for these 
forces are given in Appendix~\ref{app:3nf}.

For the 2$\pi$-exchange term, the power of this systematic scheme to 
derive consistent NN and 3N interactions comes into play. All vertices also 
occur in NN interaction diagrams. This completely determines the strength 
and form of this part of the 3NF. As outlined in the appendix, there are three 
less well known strength constants in this part of the 3NF: $c_1$, $c_3$ and $c_4$. 
They determine the contribution of various spin-isospin structures 
to the off-shell $\pi$N scattering amplitude entering the 3NF. 
Table~\ref{tab:ciconst} summarizes the results of recent determinations 
of these constants. The most fundamental determinations were done in 
Ref.~\cite{fettes98,buttiker00} using a fit to $\pi$N scattering data. 
For this system, ordinary ChPT without resummation is used. 
The possibility to extract these constants from $\pi$N data 
makes the strong link of the nuclear interaction 
and other strong interaction processes in ChPT explicit. 
The rather scarce set of data for the $\pi$N system induces rather large 
uncertainties. This has partly been circumvented in \cite{buttiker00} 
using dispersion relations. Therefore, the  determination of Ref.~\cite{buttiker00} is usually considered 
to be more accurate. Therefore, it is nice to see that 
this result compares very well with the determination of \cite{rentmeester03}, 
where the $c_i$'s were extracted from a partial wave analysis of NN data.  
For the Idaho-N3LO interaction, the authors of \cite{entem03a} preferred 
their own determination for $c_3$ and $c_4$ 
based on a fit to  phase shifts in high partial waves, 
also shown in the table. It slightly deviates from the two previous values
and lead to an intense debate on the correct way to extract 
these constants from NN data.  We cannot contribute to this discussion, but 
will use the $c_i$'s of \cite{entem03a}, simply 
because we want consistency of  the NN and 3N forces. The exact magnitude 
of these constants should be of less importance, because they only appear 
in subleading terms of minimal order $Q^3$. However, this issue has to be 
kept in mind, when we compare our results to the experimental values, especially for 
observables, where 3NF's are important.  Please note that we will combine 
an NN interaction of order $Q^4$ to a 3NF of order $Q^3$ only. Because 
we will not study the convergence of our predictions with respect to 
the chiral expansion in this work, we think that this combination 
is justified. It is the state-of-the-art for chiral interaction models. 
 
\begin{table}[t] 
\begin{tabular}{l|rrr}
                                                                  &  $c_1$    &    $c_3$ & $c_4$ \cr
\hline
Fettes et al. \cite{fettes98}                      & -1.23       &  -5.94    & 3.47 \cr
B\"uttiker et al. \cite{buttiker00}                 & -0.81       &  -4.70    & 3.40 \cr
Rentmeester et al. \cite{rentmeester03}  & -0.76      &  -4.78     & 3.96 \cr 
Entem et al. \cite{entem03a}                     & -0.81       &  -3.20    & 5.40 \cr
\end{tabular}
\caption{\label{tab:ciconst}  Various determinations of the strength constants 
$c_i$ of the 3NF. All values are in GeV$^{-1}$. }
\end{table}
    
\begin{figure}
\includegraphics[width=7cm]{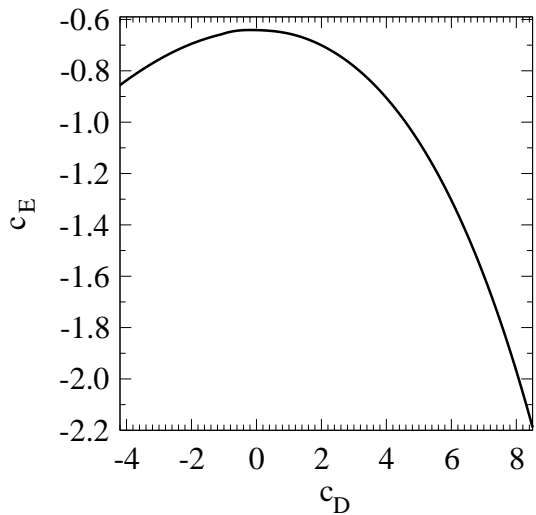}
\caption{\label{fig:cde3h} Correlation of the two dimensionless strength 
constants  $c_D$ and $c_E$ of the 3NF.  All combinations of $c_D$ and $c_E$
describe the $^3$H binding energy equally well.} 
\end{figure}

\begin{figure}
\includegraphics[width=7cm]{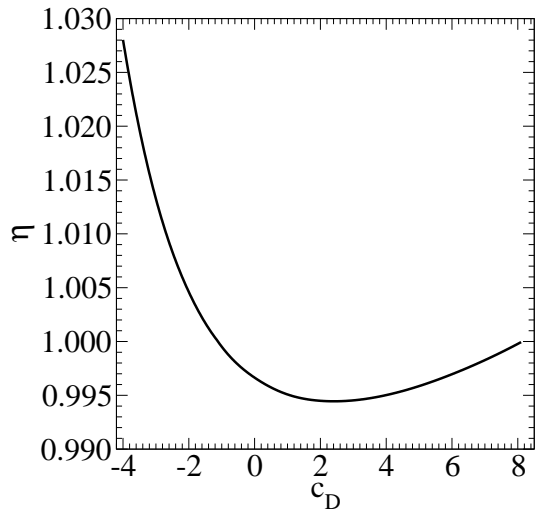}
\caption{\label{fig:cde4he} Eigenvalues $\eta$ of the Yakubovsky 
equations for $^4$He depending on $c_D$. $c_E$ is chosen according 
to Fig.~\ref{fig:cde3h}. $\eta=1$ corresponds to describing the $^4$He binding energy 
exactly.} 
\end{figure}

For the remaining two contact 3NF's, two more strength constants 
enter: $c_D$ and $c_E$ (see the appendix for the definition). In 
Ref. \cite{epelbaum02c}, it was shown that the determination of these 
constants is possible by a fit to the $^3$H binding energy and the 
Nd doublet scattering length. This is only one of the possibilities to 
determine these constants. For the Idaho-N3LO force, we therefore decided 
to fit to the $^3$H and $^4$He binding energies. For our study here, this 
has two advantages. Firstly, a very accurate fit is possible, because 
both binding energies are well known experimentally. Secondly, it makes 
a comparison of our results to calculations with the phenomenlogical model 
AV18/Urbana-IX more meaningful, because this combination describes both 
binding energies reasonably well, but fails for some $p$-shell spectra. 
The fit was performed using Faddeev-Yakubovsky calculations. For a detailed 
description of the technique, we refer to \cite{nogga02b}. 
Using the $^3$H binding energy, we established a relation between 
$c_D$ and $c_E$ shown in Fig.~\ref{fig:cde3h}. For all these combinations of $c_D$ and 
$c_E$, Idaho N3LO in combination with the 3NF predict a $^3$H binding energy 
of 8.482~MeV in agreement with experiment. These combinations then enter
in a calculation of the Yakubovsky eigenvalue $\eta$ for $^4$He using 
the $\alpha$ particle binding energy of 28.3~MeV as trial energy. Our 
theoretical prediction agrees with 28.3~MeV, if $\eta$ is equal to one. 
From the results shown in Fig.~\ref{fig:cde4he}, we can read off that two combinations 
of $c_D$ and $c_E$ describe $^3$H and $^4$He equally well.  For later 
reference, we call them ``3NF-A'' and ``3NF-B''. The numerical values 
are listed in Table~\ref{tab:cdetab}.

\begin{table}
\begin{tabular}{l|rr}
                 & $c_D$   & $c_E$ \cr
\hline
3NF-A     & -1.11        &  -0.66  \cr
3NF-B     & 8.14         & -2.02                   
\end{tabular}
\caption{\label{tab:cdetab} Combinations of  $c_D$ and $c_E$, which 
describe the $^3$H and $^4$He binding energies.}
\end{table}

We did not perform complete calculations for the binding energy for $^4$He
for each combination of $c_D$ and $c_E$. However, generally, one can 
expect from the ranges of $\eta$ results that the binding energy would change by 
approximately 2~MeV in the range of $c_D$'s considered. 
This sets a scale for the $Q^3$ contribution to $^4$He. 
For completeness, we also note that the pp force was augmented by 
the Coulomb interaction for point like protons. No further electromagnetic 
contribution was taken into account as they are  expected to be of less 
importance. We also do not include the first relativistic corrections to the 
3NF, which some people do count as an order $Q^3$ contribution. We cannot 
discuss this issue here. We only claim that, whatever counting you 
prefer, the relativistic corrections will be smaller than our $Q^3$, because we use   
cutoffs around  500~MeV, which are much smaller than the nucleon mass.  
Then, because no four-nucleon forces enter  at order $Q^3$, our interaction 
is complete up to this order.
Now we need to extend our calculations to the $p$-shell. For this, we will give 
a brief introduction to the NCSM in the next section.

\section{NCSM approach}
\label{sec:ncsm}

Usually shell-model calculations exploit the possibility to assume an
inert nuclear core. Taking only the valence nucleons as active
particles clearly has the advantage to keep the number of many-nucleon
states manageable, but, so far, it also means that effective
shell model interactions have to be used, which cannot be related 
directly to nuclear interactions, as they have been developed for
few-nucleon systems. 
The ``ab-initio no-core shell model (NCSM)'' or, more simply, 
the NCSM is different \cite{navratil00c,navratil00b}. Because all
nucleons are taken to be active, in principle, the same interactions
can be used for the NCSM as for traditional few-body calculations. 
However, for realistic nuclear forces, the wave function is hardly 
described by combinations of single particle states. Using harmonic 
oscillator (HO) states, 
which allow for an easy separation of the center of mass (CM) motion 
from the internal degrees of freedom, one encounters the additional
problem to describe the  
the exponential long range tail of the wave function. Again effective 
interactions become necessary to achieve convergence, with a small
number of HO states taken into account. These effective interactions 
are systematically related to the ``bare'' 
interactions \cite{suzuki80,suzuki82}.

The scheme described in Ref.~\cite{suzuki82} is based on unitary 
transformations of the Hamiltonian \cite{okubo54}, 
which decouples the model space from the complete Hilbert space 
describing the quantum mechanical system. 
Such transformations have already proven to be 
useful in a wide variety of nuclear physics problems. They allow us to
decouple parts of the Fock space, such as the two nucleon space, from 
the rest of the Fock space \cite{epelbaoum98b}. Also a
decoupling of low and high momentum components has been performed
using the same approach \cite{epelbaoum99,bogner03a}. In many-body problems, 
it has not only been used for shell model calculations, but, recently, 
could be exploited for the solution of nuclear problems in the hyperspherical harmonics basis \cite{barnea00,barnea01}.  
  
The starting
point of all NCSM calculations is a non-relativistic  $A$-body Hamiltonian,
which includes NN and, important here, 3N interactions
\begin{equation}
\label{eq:bareham}
H_A = \sum_{i=1}^A { {\vec { p_i}} ^2 \over 2 m_i} + \sum_{i<j=1}^A V_{ij}
+\sum_{i<j<k=1}^A V_{ijk}
\end{equation} 
$\vec p_i$ are the momenta of the $A$ particles with masses $m_i$. 
In the following, we assume that the interaction does not change 
the type of particle and that the total mass $M_A=\sum_i m_i$ is 
well defined. 
Adding the CM HO potential, one can rewrite the Hamiltonian as \cite{lipkin58}
\begin{eqnarray}
\label{eq:omham}
H_A^\Omega & = & H_A + { M_A \Omega^2 \over 2 } \ {\vec R}^2 
= \sum_{i=1}^A \left ({ {\vec {p_i}}^2 \over 2 m_i } + { 
 \ m_i \ \Omega^2 \over 2 } \ {\vec {r_i}}^2  \right)  \cr
& & +  \sum_{i<j=1}^A \left( V_{ij} - { m_i \ m_j \over 2 M_A} \ \Omega^2
 (\vec r_i - \vec r_j) ^2 \right) \cr
& & +\sum_{i<j<k=1}^A V_{ijk}
\end{eqnarray}
with the CM position
$\vec R = { 1 \over M_A } \ \sum_i m_i \vec r_i$.

In the following we would like to establish an unitary tranformation
of the Hamiltonian, which decouples two parts of the Hilbert space. A
rather small finite model space $P$ and the rest of the Hilbert space
$Q$. The projection operators on the two spaces are also called $P$ and
$Q$. They fulfill the relation $Q=1-P$. 

Okubo \cite{okubo54} showed that the unitary operator can be 
derived from an operator $\omega$, which
fulfills the following relations:
\begin{eqnarray}
\label{eq:omprop1}
\omega & =  & Q \ \omega \ P \\
\label{eq:omprop2}
0 & = & Q \ e^{-\omega} \ H_A^\Omega \ e^{+\omega} \ P \cr 
  & = & Q \ (1-\omega) \ H_A^\Omega \ (1+\omega) \ P  
\end{eqnarray}
The transformation $e^{\omega}$ already decouples the subspaces $P$
and $Q$, but it is not unitary. Unitarity can be achieved using 
the antihermitian operator $S={\rm arctanh }(\omega - \omega^\dagger)$ 
and defining the unitary operator $U=e^{+S}$. Using the properties 
of $S$ and $\omega$, one finds the explicit expression \cite{okubo54}
\begin{equation}
U = e^{S} = (1+\omega-\omega^\dagger) \ ( 1 + \omega \ \omega^\dagger
+ \omega^\dagger \ \omega ) ^ { -1/2} 
\end{equation}
The problem is reduced to finding the operator $\omega$. For that, one
defines a set of $d_P$ eigenstates $| n \rangle$ of the Hamilton operator 
$H_A^\Omega$, where $d_P$ is the dimension of the model space $P$.
Defining $ \omega   \ | n \rangle \equiv  Q \ | n \rangle$, it is 
easy to show that the decoupling condition Eq.~(\ref{eq:omprop2}) 
holds. This completely defines the $\omega$ operator for sets of  
eigenstates $| n \rangle$, for which the $P \ | n \rangle$ are linearly 
independent. In practice, for resonable choices of the eigenstates,
this condition is always fulfilled. Using Eq.~(\ref{eq:omprop1}) the action of $\omega$ on
$|n \rangle $ reads 
\begin{eqnarray}
 \langle \alpha_Q | n  \rangle  & = & \sum_{\alpha_P}  \langle  \alpha_Q | \ \omega \
| \alpha_P  \rangle   \langle  \alpha_P | n  \rangle  \cr
& \equiv &  
\sum_{\alpha_P}  \langle  \alpha_Q | \ \omega \
| \alpha_P  \rangle  N_{\alpha_P n}
\end{eqnarray}
$\alpha$ enumerates the basis states. The indices $P$ and $Q$ 
restrict enumeration to the respective subspace. 
With the help of the inverse $M=N^{-1}$, the matrix 
elements of $\omega$ are found  \cite{navratil96} easily in a non-iterative
scheme
\begin{equation}
 \langle  \alpha_Q | \ \omega \ | \alpha_P  \rangle  = \sum_n  \langle  \alpha _Q | n  \rangle  
\ M_{n\ \alpha_P} 
\end{equation}
As elaborated in \cite{viazminsky01}, there is always 
as set of states $|n \rangle $, for which the inverse of $N$ exists. 
In practice, one chooses the $d_P$ lowest lying eigenstates of
$H_A^\Omega$. Because $(1-\omega+\omega^\dagger) | n \rangle $ has only 
$P$-space components, the same holds for $U^\dagger | n \rangle $. 
Matrix elements of states $|n \rangle $ are then exactly
reproduced in the model space
\begin{equation}
 \langle  n' | \ O \ |  n  \rangle   =  \langle  n ' | U \
\underbrace {P U^\dagger \ O \ U P }_{O_{eff}} \ U^\dagger | n  \rangle   
\end{equation}
Because we are only interested in the $P$-space matrix 
elements of $O_{eff}$, we can make use of the simplification 
\begin{equation}
U P = (1+\omega) \ ( 1 + \omega^\dagger \ \omega ) ^ { -1/2} P
\end{equation}
and write 
\begin{eqnarray}
\label{eq:effop}
& &  \langle  \alpha_P | \ O_{eff} \ | \beta_P  \rangle  
= \sum_{\alpha_P ' \ \beta_P '}
 \langle  \alpha_P | ( 1 + \omega^\dagger \ \omega ) ^ { -1/2} | \alpha_P '  \rangle  \cr
& & \qquad \sum_{n \ n'} M^*_{n \alpha_P'} M_{n' \beta_P'} \ 
 \langle  n | \ O \ | n '  \rangle   \cr
& & \qquad \qquad 
 \langle  \beta_P ' | ( 1 + \omega^\dagger \ \omega ) ^ { -1/2} | \beta_P   \rangle 
\end{eqnarray}
Here we made use of the relation
\begin{eqnarray}
\label{eq:nonhermeff}
& &  \langle  \alpha_P | (1+\omega^\dagger)\  O \ (1+\omega) | \beta_P  \rangle  \cr
& & \quad = 
\sum_{n \ n'} M^*_{n \alpha_P} M_{n' \beta_P} \ 
 \langle  n | \ O \ | n '  \rangle   
\end{eqnarray}
There are two special cases of interest of Eq.~(\ref{eq:nonhermeff}).
The first, $O=1$, is useful to determine the $P$ space renormalization 
operator $( 1 + \omega^\dagger \ \omega ) ^ { -1/2}$ since
\begin{eqnarray}
& &  \langle  \alpha_P | (1+\omega^\dagger) \ (1+\omega) | \beta_P  \rangle  \cr 
& & \quad =  
 \langle  \alpha_P | (1+\omega^\dagger\omega) | \beta_P  \rangle  = 
\sum_{n} M^*_{n \alpha_P} M_{n \beta_P} 
\end{eqnarray}
The second, $O=H_A^\Omega$, makes the evaluation of the effective 
Hamiltonian especially simple
\begin{eqnarray}
 \langle  \alpha_P |  {\cal H}_{eff} | \beta_P  \rangle &  = & \langle  \alpha_P | (1+\omega^\dagger)\  H_A^\Omega \ (1+\omega) | \beta_P  \rangle  \cr
 & & = \sum_{n} M^*_{n \alpha_P} M_{n \beta_P} \ E_n
\end{eqnarray}
with $E_n$ the eigenenergies of the eigenstates $|n \rangle $. 
These equations are exact. They reduce the problem to one in 
a very small model space $P$. However, because solutions of the
problem are necessary to obtain the effective operator, they do not
help to solve the eigenvalue problem in this form.  

The NCSM concept becomes useful by approximating the effective interaction 
in an $a$-body cluster approximation. There are two obvious constraints on this approach. 
Firstly, to keep the approximation controllable, we require that the effective interaction 
approaches the bare interaction for model spaces $P \to \infty$, and, secondly, 
we expect best convergence, when the nuclear part of the interaction acts in the $a$-body cluster  with the same strength as in a free $a$-body system. Then, under the assumption that only $b=2$-body bare forces act, one needs to solve the 
cluster problem for  
\begin{eqnarray}
\label{eq:2cluham}
H_a^\Omega & = & \sum_{i=1}^a \left ({ {\vec {p_i}}^2 \over 2 m_i } + { 
 \ m_i \ \Omega^2 \over 2 } \ {\vec {r_i}}^2  \right)  \cr 
 & & +  
 \sum_{i<j=1}^a \left( V_{ij} - { m_i \ m_j \over 2 M_A} \ \Omega^2
 (\vec r_i - \vec r_j) ^2 \right) 
\end{eqnarray}
Note that in this cluster Hamiltonian, the strength of the HO two-body potential 
depends on the original $A$. This induces a confining meanfield potential 
for the cluster, which simply reflects that the cluster is embedded in the 
nuclear environment. It has the nice property that we can base our effective 
interaction on an infinite set of cluster bound states. No scattering states need 
to be considered. The effective interaction for $b$-body interactions in the $a$-body 
cluster approximation for the $A$-body system is defined as 
\begin{equation}
{\cal V}_{eff}^{A,a,b} = {\cal H}_{eff} - \sum_{i=1}^a \left ({ {\vec {p_i}}^2 \over 2 m_i } + { 
 \ m_i \ \Omega^2 \over 2 } \ {\vec {r_i}}^2  \right) 
\end{equation}
where ${\cal H}_{eff}$ is derived from an Hamiltonian operator like given in 
Eq.~(\ref{eq:2cluham}). The effective interaction enters into the $A$-body problem 
with a different weight, that insures that the bare interaction is recovered for 
large model spaces \cite{navratil00c,navratil00b}
\begin{eqnarray}
{\cal H}^A_{eff} & = & \sum_{i=1}^A \left ({ {\vec {p_i}}^2 \over 2 m_i } + { 
 \ m_i \ \Omega^2 \over 2 } \ {\vec {r_i}}^2  \right)   \cr 
  & &  + {  (A-a)! (a-b)!  \over (A-b)! } \ 
   \sum_{i_1 < i_2...< i_a=1}^A {\cal V}_{eff,i_1 ... i_a}^{A,a,b}  
\end{eqnarray}
In this form it becomes apparent that we have a double counting problem, 
when both, $b=2$ and $b=3$ interactions enter the cluster Hamiltonian. 
This can be solved as outlined in \cite{navratil03} and requires two effective 
interactions. The first one, ${\cal V}_{eff,i_1 ... i_a}^{A,a,NN}$,   
is based on the Hamiltonian in Eq.~(\ref{eq:2cluham}). The second one,
${\cal V}_{eff,i_1 ... i_a}^{A,a,3N}$, is based on 
\begin{eqnarray}
\label{eq:3cluham}
H_a^\Omega & = & \sum_{i=1}^a \left ({ {\vec {p_i}}^2 \over 2 m_i } + { 
 \ m_i \ \Omega^2 \over 2 } \ {\vec {r_i}}^2  \right)  \cr 
 & & +  
 \sum_{i<j=1}^a \left( V_{ij} - { m_i \ m_j \over 2 M_A} \ \Omega^2
 (\vec r_i - \vec r_j) ^2 \right) \cr
& & + \sum_{i<j<k=1}^a V_{ijk}
\end{eqnarray}
which additionally takes into account the 3NF. The combination 
\begin{eqnarray}
{\cal H}^A_{eff} & = & \sum_{i=1}^A \left ({ {\vec {p_i}}^2 \over 2 m_i } + { 
 \ m_i \ \Omega^2 \over 2 } \ {\vec {r_i}}^2  \right)   \cr 
  & &  + {   (A-a)!  (a-3)! \over (A-3)!   } \ 
   \sum_{i_1 < ... < i_a=1}^A {\cal V}_{eff,i_1 ... i_a}^{A,a,3N}  \cr
   & &  + {   (A-a)! ((A-3)! (a-2)!  -(A-2)! (a-3)!  )  \over (A-3)!   (A-2)!  } \ 
   \cr & & \qquad \sum_{i_1 < ... < i_a=1}^A {\cal V}_{eff,i_1 ... i_a}^{A,a,NN}
\end{eqnarray}
then fulfills the constraints for our effective interaction.
Specifically for $a=3$, we get  
\begin{eqnarray}
{\cal H}^A_{eff}
   & =  & \sum_{i=1}^A \left ({ {\vec {p_i}}^2 \over 2 m_i }  
   + {  \ m_i \ \Omega^2 \over 2 } \ {\vec {r_i}}^2  \right)   \cr 
  & &  + \sum_{i_1 < ... < i_a=1}^A {\cal V}_{eff,i_1 ... i_a}^{A,a,3N} 
  \cr & & 
  + {   3-A   \over A-2  }  \sum_{i_1 < ... < i_a=1}^A {\cal V}_{eff,i_1 ... i_a}^{A,a,NN}  \end{eqnarray}   
This interaction will be used in the calculations in the next section.
It will enhance the convergence so that we can obtain meaningful binding 
energies for $p$-shell nuclei. In the following, we also use the $a=3$ approximation,
when only ``bare'' NN forces are taken into account, i.e., when $V_{ijk}=0$.

\section{Results}
\label{sec:results}

The main subject of this section will be a careful analysis of the 
numerical accuracy and convergence properties of the results.
The cluster approximation induces a sizeable $\Omega$ 
dependence, which is driven by the mismatch of the long range 
behavior of the cluster states and the wave function for the 
nucleus. The former ones being confined by a HO behave like 
Gaussians, whereas the latter ones decay exponentially for large 
distances. Because renormalization of the operators 
is generally weak for the large distances \cite{stetcu05a}, this cannot 
be corrected by the renormalization of the interaction, but needs 
to be taken care of by an optimization of the $\Omega$ for the problem 
and by a rather large model space, which includes enough 
basis states that the tail is also well approximated.   
Therefore, to get reliable results, calculations for 
different model spaces have to be performed for a range of $\Omega$'s
to identify the optimal choice \cite{navratil04}.

The model space size is characterized by the maximal 
number $N$ of HO excitations included. The lowest $^7$Li 
HO configuration has four nucleons in the $n=0$ HO states and 
three in the $n=1$ shell. Therefore, the
 minimal HO energy is ${\cal N}_{min}=3$. The $A$-body calculations include 
 states up to ${\cal N}_{max} =   {\cal N}_{min} + N$ \footnote{Note that in 
 previous works, what we call ``$N$'' here, was labeled ``$N_{max}$''.}. 
 In this work, we will perform 
 calculations up to $N=6$ including configurations up to ${\cal N}_{max}=9$. 
 The calculations were performed in a basis of single particle 
 Slater-deterimants ($m$-scheme) using the Many-Fermion-Dynamics 
 code \cite{varymfd1,varymfd2}. Since we employ HO single particle states 
 and all Slater determinants up to ${\cal N}_{max}$ are included in the model 
 space, the center of mass part can be exactly separated at the end. 
 For $N=6$, the P-space dimension was 663527
requiring  approximately $3 \cdot 10 ^9$ matrix elements for the $A$-body Hamiltonian. 
The effective interactions are based on $a=3$ clusters. For each model space, 
we include three-body states up to the maximal excitation of three particles 
in the $A$-body configurations. For our largest calculation, these are three-body 
states up to ${\cal N}_3 = 9$ and, for the smaller calculations, this number decreases 
correspondingly. This implies that the effective interactions are different 
for each model space size. The results for smaller model spaces do not possess  
upper bound character. 

To solve the cluster equation,
we use the Lanczos-scheme. The usage of a Jacobi basis for this problem 
guarantees the translational invariance of the effective interaction. This is 
also mandatory to exclude unphysical center of mass excitations from 
the model space, which would also complicate the selection of the 
proper states $| n \rangle$ for the definition of the $\omega$ operator. 
The cluster equations are solved using the ``bare'' interactions 
for Q-spaces up to ${\cal N}_3=48$ for three-body angular momenta 
up to $J_3= { 7 \over 2 }$ and ${\cal N}_3=40$ for $J_3= { 9 \over 2 }$
to $21 \over 2$. The largest dimension for the  ${\cal N}_3 = 9$ 
P-space is 157 for $J_3^\pi = {5 \over 2}^-$ and three-body isospin 
$T_3 = {1 \over 2}$ making the same amount of 
bound state solutions necessary in this channel.  Altogether, for 
all channels between $J_3= { 1 \over 2 }$  and $J_3= { 21 \over 2 }$, we 
required 2352 states.  Convergences for the cluster states 
has been checked using their binding energies. We take 
the charge dependence of the NN interaction into account by averaging 
the NN isospin $t=1$ interactions according to the prescription 
given in \cite{kamuntavicius99}. This averaging was performed for 
total isospin $T={1/2}$ $^7$Li states.   All states we will discuss here 
are $T=1/2$ states. The Coulomb interaction is not included 
into the calculation of the effective interaction, but is taken into account 
as an additional ``bare'' force during the solution of the $A$-body 
problem. This does not change any  results, but will facilitate 
calculations for $T=3/2$ states later, which can now be well 
approximated using the same effective interactions.

\begin{figure}
\includegraphics[width=7cm]{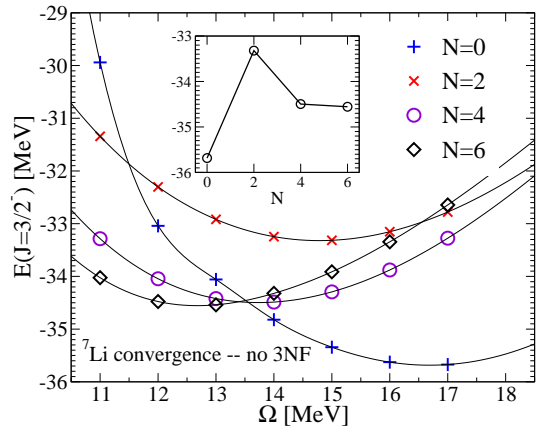}
\caption{\label{fig:li7no3nfgsom} (Color online) 
$\Omega$ dependence of the binding energy for the $^7$Li $J^\pi ={3 \over 2}^-$ 
ground state using the NN interaction only. Results for $N=0$ to $N=6$ model spaces are shown. }
\end{figure}

\begin{figure}
\includegraphics[width=7cm]{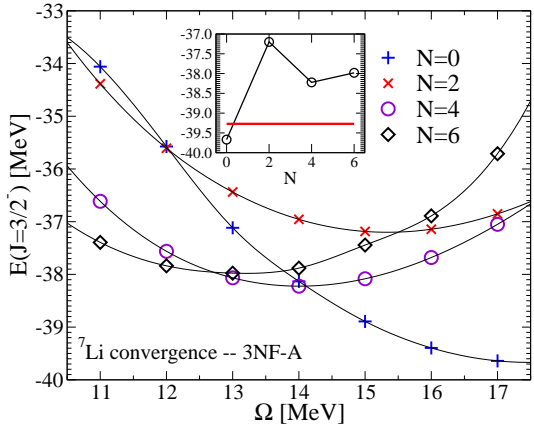}
\caption{\label{fig:li73nfagsom} (Color online) 
$\Omega$ dependence of the binding energy for the $^7$Li $J^\pi ={3 \over 2}^-$ 
ground state using the NN interaction combined with 3NF-A. 
Results for $N=0$ to $N=6$ model spaces are shown. }
\end{figure}

\begin{figure}
\includegraphics[width=7cm]{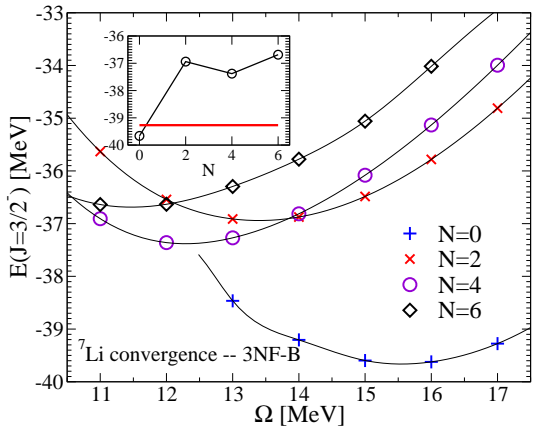}
\caption{\label{fig:li73nfbgsom} (Color online) 
$\Omega$ dependence of the binding energy for the $^7$Li $J^\pi ={3 \over 2}^-$ 
ground state using the NN interaction in 3NF-B. 
Results for $N=0$ to $N=6$ model spaces are shown. }
\end{figure}

In Figs.~\ref{fig:li7no3nfgsom}, \ref{fig:li73nfagsom} and \ref{fig:li73nfbgsom}, 
we show the $\Omega$ dependence of the 
ground state binding energy for $^7$Li with the Idaho-N3LO interaction
only and the combinations with 3NF-A and 3NF-B, respectively. 
The binding energy exhibits a distinct minimum for all 
model spaces and interactions shown. The position of this minimum shifts 
to smaller $\Omega$ for larger model spaces. This shift becomes smaller 
for the larger spaces. Also, the position depends on the 
interaction model. This behavior has been observed for all realistic 
interactions and makes the extraction of binding energy results more 
difficult than for the simpler Minnesota type of interactions. 
Ref.~\cite{navratil04} investigates 
this issue in detail. There, it was shown that taking 
the binding energy of the minimum for different model spaces 
results in a clear pattern of convergence. It was also shown that 
the results obtained in this  way are consistent with results based on the ``bare'' interaction. Therefore, we also follow this procedure. The inserts in the figures  
show the convergence of the binding energy 
based on this idea for model spaces with $N=0$, 2, 4 and 6. Very clearly we observe
convergence for Idaho-N3LO without 3NF and with 3NF-A. For 3NF-B, 
the changes in the step $N=4$ to $N=6$ are somewhat larger. 
They are, however, small enough  to justify that the results are accurate to 
the order of the few hundred keV's or one MeV.
This will be sufficient for a discussion of the  quality of chiral interactions later. 

\begin{figure}
\includegraphics[width=7cm]{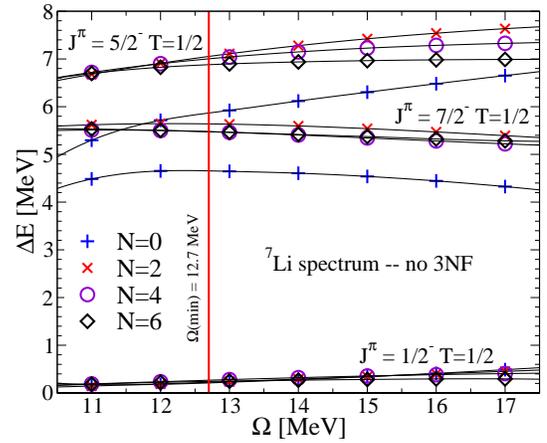}
\caption{\label{fig:li7no3nfspectrom} (Color online) 
$\Omega$ dependence of the excitation energy of the lowest 
states of $^7$Li using the NN interaction only. 
Results for $N=0$ to $N=6$ model spaces are shown. }
\end{figure}

\begin{figure}
\includegraphics[width=7cm]{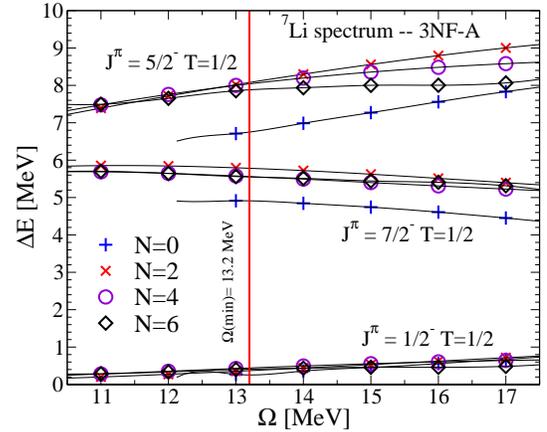}
\caption{\label{fig:li73nfaspectrom} (Color online) 
$\Omega$ dependence of the excitation energy of the lowest 
states of $^7$Li using the NN interaction combined with 3NF-A. 
Results for $N=0$ to $N=6$ model spaces are shown. }
\end{figure}

\begin{figure}
\includegraphics[width=7cm]{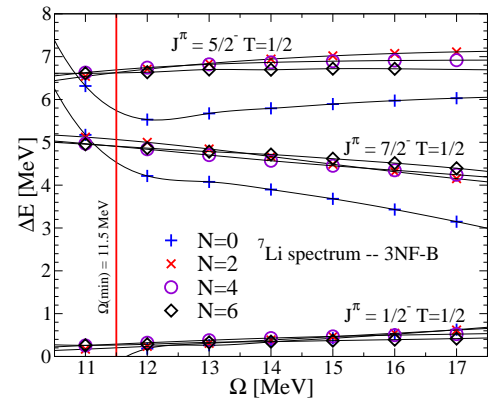}
\caption{\label{fig:li73nfbspectrom} (Color online) 
$\Omega$ dependence of the excitation energy of the lowest 
states of $^7$Li using the NN interaction combined with 3NF-B. 
Results for $N=0$ to $N=6$ model spaces are shown. }
\end{figure}

The results for the excitation energies are shown in 
Figs.~\ref{fig:li7no3nfspectrom} to \ref{fig:li73nfbspectrom}.
Fortunately, the $\Omega$ dependence is rather 
mild and the different excitation energies are very similar 
for the $N=2$, 4 and 6 model spaces. The vertical lines indicate 
the $\Omega$ of the minima for $N=6$ in Figs.~\ref{fig:li7no3nfgsom} to 
\ref{fig:li73nfbgsom}. Our final extraction for the excitation energies 
is the $N=6$ result at this $\Omega$ value. We also  checked the $\Omega$ 
dependence of the binding energy for the excited $1/2^-$ state. Since the position 
of the minima is very similar to the ones for the $3/2^-$ state, 
we are confident that the excitation energies for the  shown 
$3/2^-$, $1/2^-$, $7/2^-$ and $5/2^-$ are accurate.

Our results for the binding energy are summarized and compared 
to other calculations and the experiment in Table~\ref{tab:li7bind}.
Besides our new results, we show results for phenomenological interactions 
from GFMC calculations. 
An earlier NCSM results for AV8' and TM' \cite{navratil03} 
gave 35.8~MeV. It was based only 
on a single value of   $\hbar \Omega$, i.e. $14$~MeV, and on 
three-body cluster states  up to ${\cal N}_3 = 28$.
For a more reliable binding energy prediction, a complete study 
of the $\Omega$ dependence is certainly required as well as 
an increased number of  three-body cluster states. Preliminary results 
clearly show this trend. Note that we use  ${\cal N}_3 = 48$ in the present study. 
Because of this, we omit the earlier NCSM result 
here and compare to the GFMC results. 

As expected, $^7$Li is underbound for the NN interaction only. Both  
3NF's  provide more binding. However, in both cases 
the final binding energy result is still short of the experiment by 1.2 and 2.5~MeV,
respectively. In view of the general expectation that strong 
repulsion at short distances is required to avoid a collapse of nuclei,
we are encouraged by the slight underbinding. Clearly, we do not observe 
any sign of overbinding, though neither the NN force nor the 3NF have a 
strong repulsive core. Both interactions are very soft, yet the binding energies 
are reasonable. For phenomenogical, pure 2$\pi$ exchange 3NF's, like 
the Tucson-Melbourne (TM') \cite{coon01}, overbinding already sets in 
for $^7$Li.  So far, it was believed that only the addition of 
a repulsive core, like in the Urbana-IX and Illinois models 
\cite{pudliner97,pieper01a}, can cure this overbinding problem. 
Here, we show that the additional structures of chiral 3NF's also prevent 
overbinding. We note that the description of the binding energy is best for 
3NF-A.

\begin{table}
\begin{tabular}{l|rr}
model    &    $E_{gs}$ [MeV] & $r$ [fm] \cr
\hline 
NN only     &  34.6  &  2.15  \cr
w/ 3NF-A   &  38.0  &  2.11 \cr 
w/ 3NF-B   &  36.7  &  2.23 \cr
\hline
AV8'+TM' \cite{pieper05a,pieperpriv}           &   40.5(1) & 2.2(1)  \cr 
AV18+Urbana IX \cite{pieper04,pieper01a}  &   37.5     & 2.33 \cr 
AV18+IL2            \cite{pieper04,pieper01a}  &   38.9     & 2.25 \cr 
\hline
Expt.  \cite{tilley02,vries87}                   &   39.2    & 2.27 \cr 
\end{tabular}
\caption{\label{tab:li7bind} Comparison of the ground state binding 
energy  results $E_{gs}$ and the point proton 
rms radius $r$ for $^7$Li (for $N=6$ model space and $\Omega$ 
at the minimum of the energy/$\Omega$ curve) 
for  chiral interactions and several phenomenological 
combinations to the experimental value. The experimental rms radius
is corrected for the finite size of the proton assuming 
a proton charge radius of 0.81~fm$^{-1}$. }
\end{table}

\begin{figure}
\includegraphics[width=7cm]{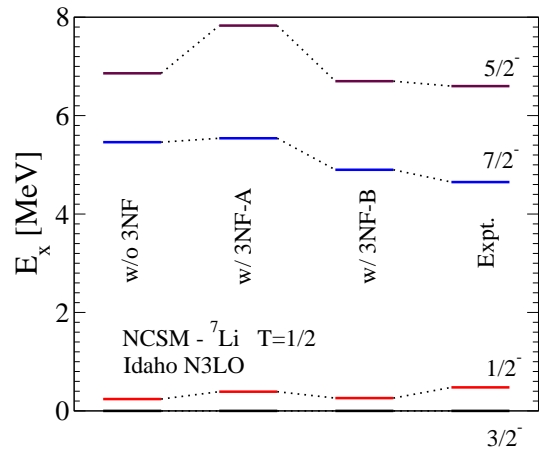}
\caption{\label{fig:li7no3nfdepspectr} (Color online) 
Dependence of the excitation energy of the lowest 
states of $^7$Li on the interaction. Results with the NN interaction only and
with the 3NF-A and 3NF-B included are compared to the experimental value. }
\end{figure}

The final results for the excitation energies are summarized in 
Fig.~\ref{fig:li7no3nfdepspectr}. All combinations of the  
interactions, Idaho-N3LO alone, with 3NF-A or 3NF-B, 
do predict the right ordering for these states. The splitting of the 
$3/2^-$ and $1/2^-$ states is small. The agreement with
this experimental splitting seems to be superior for 3NF-A. Because the 
splitting itself is very small, this might be accidental. More 
significant deviations of the predictions are observed for 
the $7/2^-$ and $5/2^-$ states. Both, the position of this 
multiplet  and the splitting is strongly affected by the 3NF's
and the agreement with the experimental results 
is clearly best for 3NF-B. This is in contradiction to the binding 
energy, which was best described for parameter set 3NF-A.
For a better clarification of this situation, a study of  the 
dependence on the choice of the $c_i$'s is mandatory. 
This issue is not as simple as it seems at  first glance, since the same 
$c_i$'s enter the NN interaction. So, in principle, a refitting 
of the NN interaction would be necessary. We postponed this issue
to a later study, where this constraint might be loosened 
to get more insight into the $c_i$ dependence.    
Finally, we note that part of the deviations for the binding 
energy might be attributed to our numerically uncertainty. 
Therefore, the results for $^7$Li favor 
the parameter set 3NF-B, because of its superior description of 
the spectrum, but further calculations will be required before we can conclude 
that this result also holds for heavier nuclei.

\begin{figure}
\includegraphics[width=7cm]{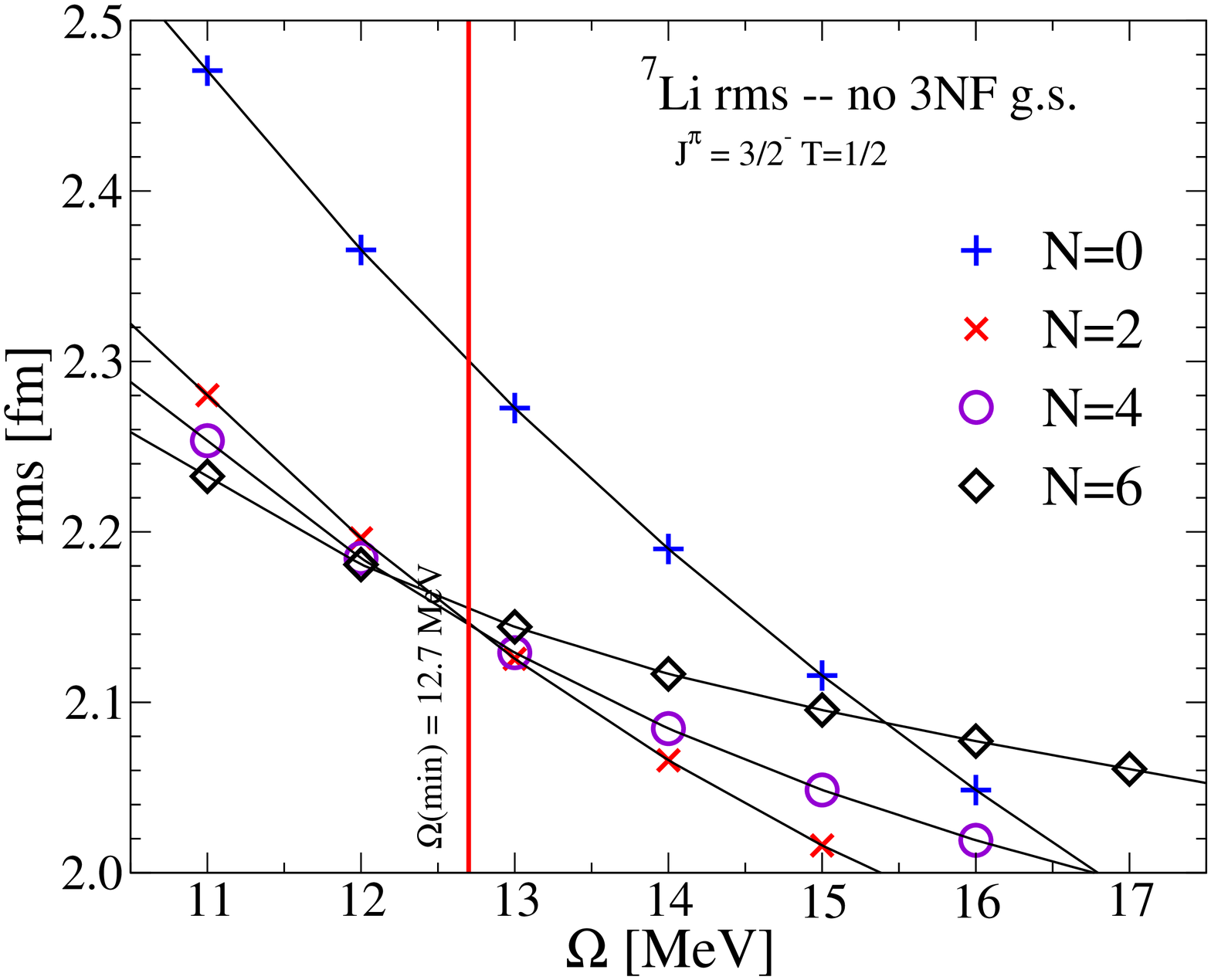}
\caption{\label{fig:li7no3nfrmsom} (Color online) 
$\Omega$ dependence of the point proton 
rms radius of the ground state 
of $^7$Li using the NN interaction only.
Results for $N=0$ to $N=6$ model spaces are shown. }
\end{figure}

\begin{figure}
\includegraphics[width=7cm]{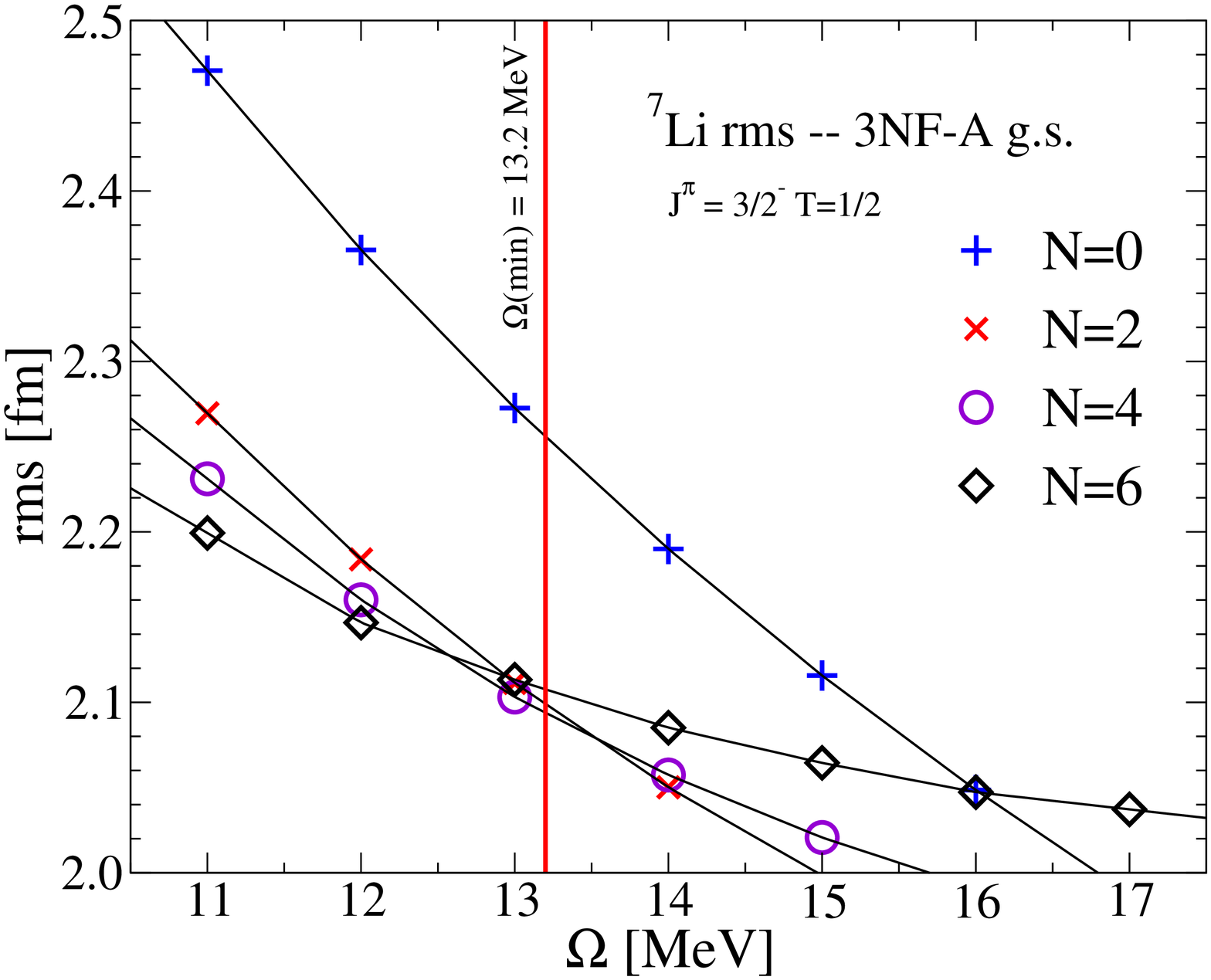}
\caption{\label{fig:li73nfarmsom} (Color online) 
$\Omega$ dependence of the point proton 
rms radius of the ground state 
of $^7$Li using the NN interaction combined with 3NF-A.
Results for $N=0$ to $N=6$ model spaces are shown. }
\end{figure}

\begin{figure}
\includegraphics[width=7cm]{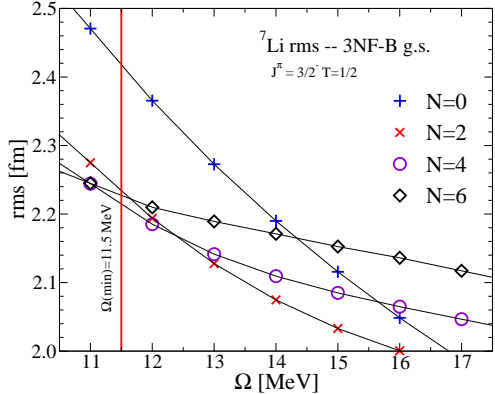}
\caption{\label{fig:li73nfbrmsom} (Color online) 
$\Omega$ dependence of the point proton 
rms radius of the ground state 
of $^7$Li using the NN interaction combined with 3NF-B.
Results for $N=0$ to $N=6$ model spaces are shown. }
\end{figure}

As we discussed before, we consider one main result of this first 
calculation to be that the very soft chiral interactions do not predict 
a collapse of nuclei with increasing number of nucleons. Of course, 
this also needs to be confirmed for the densities. Therefore, we also 
show results for the point proton radii for the $3/2^-$ state in 
Figs.~\ref{fig:li7no3nfrmsom} to \ref{fig:li73nfbrmsom}. 
For this quantity, we found an interesting dependence 
on $\Omega$ and $N$. For all model spaces, the radii decrease with 
increasing $\Omega$. This exactly reflects  the behavior of the basis 
HO states. Obviously, there is no correlation to the binding energy 
result. It is comforting to note that the slope of this dependence 
decreases with increasing $N$. Also, we observed 
the smallest dependence on $N$ for the $\Omega$ 
values around the minima of the binding energy curves, which 
are again indicated by vertical lines. Therefore, we have good
indications that the best extraction of the radii are obtained for 
exactly these $\Omega$ values for the largest possible model spaces. 

Table~\ref{tab:li7bind} also shows the radii obtained in this way
and compares them to the experimental values and other 
calculations.  We obtain very realistic results for the chiral  interactions, which 
are comparable to the phenomenological models Urbana-IX, Illinois and TM'. 
We do not find any indication that soft, chiral 
interactions fail to saturate nuclear systems with a realistic binding energy 
and density.  We note that for 3NF-B we even observe an increase 
of the radius though it provides additional attraction. 
 
\section{Conclusions}
\label{sec:concl}

We have presented the first microscopic 
calculation of the low lying  $^7$Li states based on 
a full chiral interaction, including NN and 3N forces. Our results 
are based on the NCSM. We showed that predictions are possible 
with an accuracy that allows one to discuss the basic 
properties of the nuclear force. The results confirm that binding energies 
and spectra depend on the structure of the 3NF's. This means that the 
results for different sets of parameters for the 3NF are different 
though both models describe the $^3$H and $^4$He nuclei with the same 
accuracy. In this first application, we restricted ourselves to changes in
the strength of the short range, or short-range long range pieces (D- and E-terms)
of the 3NF, which are not related to the NN interaction at the order 
$Q^3$.
With this constraint,  we have already obtained reasonable 
results for both, the binding energies and the spectra, but we could 
not describe both simultaneously 
within the accuracy, which we expect from our many-body  method.  

On the other hand, we know that  the strength of 2$\pi$ exchange 
in the NN and 3N interactions
is not well determined yet. Therefore, it is necessary to establish a measure 
for the dependence of the $^7$Li and other $p$-shell results 
on this strength. Since, in principle, this can only be done by changing 
the NN and 3N forces, it is not a straightforward and simple 
task and, probably, for first estimates, one will need to loosen the strict consistency 
of NN and 3N forces somewhat. 
This will be the subject of a forthcoming study.

Here, we were mainly interested to establish that soft chiral interactions 
result in realistic binding energies and densities. We therefore also 
investigated the point proton radii and described how these can be extracted 
from NCSM wave functions. The results support that soft, low momentum 
interactions predict reasonable binding energies and densities. 
In the future, this conjecture needs to be confirmed for heavier systems. 
In this respect, a study of $^{10}$B is a high priority. Results employing 
the TM' 3NF and the AV8' NN force indicate an increasing 
overbinding going from $A=7$ to $A=10$ \cite{pieper05a}. 
It needs to be confirmed that the chiral interactions 
can again avoid overbinding. At the same time, $^{10}$B has proven 
to be very sensitive to the spin-orbit interactions including 
the 3NF's. Usual  NN interactions, like Idaho-N3LO 
and AV18 \cite{navratil03,navratil04,pieper01a},  
predict a wrong ordering of the  $3^+$ ground and $1^+$ excited state. 
The Illinois type of 3NF's correctly change this ordering and also the TM'
3NF affects this positively, in contrast to the Urbana IX 3NF. 
Therefore, the predictions of the complete chiral interaction with the 
3NF are especially interesting for this nucleus. But also for this system,
the dependence on the $c_i$'s of the 2$\pi$ exchange should be studied.   

\begin{acknowledgments}

We acknowledge partial support by the US DOE under Grants No.
DE-FC02-01ER41187, DE-FG02-00ER41132, and DE-FG-02-87ER-40371,
and by NSF Grant Nos. PHY0070858 and PHY0244389. 
This work was partly performed under the auspices of the U.S. Department
of Energy by the University of California, Lawrence Livermore
National Laboratory under Contract No.W-7405-Eng-48. 
Support from the LDRD Contract No. 04-ERD-058, and from the U.S. 
Department of Energy, Office of Science, (Work Proposal Number SCW0498) is 
acknowledged. A.N. and B.R.B were partially supported by 
the NATO Collaborative Linkage Grant 2004/2005.   
The numerical calculations have been performed on Seaborg 
at the National Energy Research Scientific Computing Center (NERSC), 
Berkeley. 

\end{acknowledgments}

\appendix

\section{Explicit expressions for the chiral 3NF's}
\label{app:3nf}

In  this appendix, we would like to summarize the explicit form 
of the chiral 3NF. The explicit expressions were first derived in 
\cite{kolck94}. The 3NF consists of terms with the three different 
topologies. The first one is the 
usual $2\pi$ exchange. The operator structure is identical to the 
well known Tucson-Melbourne interaction \cite{coon79}. Then there 
are two new topologies, which contain also contact interactions
(see e.g. \cite{epelbaum02c}). 
In the original work, five independent terms with such 
structures  are derived. In \cite{epelbaum02c}, it was realized that 
the number of independent terms is smaller once the antisymmetry of 
the 3N states is taken into account. Then only one term of the $1\pi$ 
exchange type (D-term) and one term of the pure contact type (E-term) 
are independent. 

Following the notation of \cite{friar99a}, the $2\pi$ exchange part 
reads
\begin{eqnarray}
\label{3nf2pi}
V_{ijk}^{(k); 2\pi}=\sum_{i \not= j \not= k} \frac{1}{2}\left(
  \frac{g_A}{2 F_\pi} \right)^2 \frac{( \vec \sigma_i \cdot \vec q_{i}) 
(\vec \sigma_j \cdot \vec q_j  )}{(\vec q_i\, ^2 + m_\pi^2 ) ( \vec
q_j\, ^2 + m_\pi^2)}  F^{\alpha \beta}_{ijk} \tau_i^\alpha 
\tau_j^\beta \cr
\end{eqnarray}
where  $\vec q_i$ is the momentum of the pion exchanged 
between nucleons $i$ and $k$ and 
\begin{equation}
F^{\alpha \beta}_{ijk} = \delta^{\alpha \beta} \left[ - \frac{4 c_1
    m_\pi^2}{F_\pi^2}  + \frac{2 c_3}{F_\pi^2}  
\vec q_i \cdot \vec q_j \right] + \sum_{\gamma} \frac{c_4}{F_\pi^2} \epsilon^{\alpha
\beta \gamma} \tau_k^\gamma  
\vec \sigma_k \cdot [ \vec q_i \times \vec q_j  ]
\end{equation}
The strength of these terms is completely determined by the 
$c_i$ constants discussed in the main text. 
The two new terms read 
\begin{eqnarray}
\label{3nf1pi}
V_{ijk}^{(k);1\pi} &=& - \sum_{i \not= j \not= k} \frac{g_A}{8
  F_\pi^2} \, \frac{c_D}{F_\pi^2 \Lambda_\chi}  \, \frac{\vec \sigma_j \cdot \vec q_j }{\vec q_j\, ^2
  + m_\pi^2}  
\, \left( {\bf \tau}_i \cdot  {\bf \tau}_j \right) 
(\vec \sigma_i \cdot \vec q_j ) \cr
\end{eqnarray}
and 
\begin{eqnarray}
\label{3nfcont}
V_{ijk}^{(k);\rm cont} &=& \frac{1}{2} \sum_{i \not= j \not= k}  \frac{c_E}{F_\pi^4 \Lambda_\chi} \,
 ( {\bf \tau}_j \cdot {\bf \tau}_k ) 
\end{eqnarray}
In these terms two new strength constants $c_D$ and $c_E$ appear. 
The values, which we determine as described in the main text from the 
binding energies of $^3$H and $^4$He, are based on the choice  
$\Lambda_\chi =700$~MeV. For the axial vector coupling constant,
we use $g_A =1.29$. $F_\pi = 92.4$~MeV is the weak pion decay 
constant and  $m_\pi=138.03$~MeV the averaged pion mass. 
As outlined in \cite{epelbaum02c}, we regularize the 3NF using a 
cutoff. Throughout this paper, we fix the cutoff to $\Lambda=500$~MeV. 

\bibliography{lit101105}

\end{document}